\documentclass[aps,pra,twocolumn,superscriptaddress]{revtex4-2}

\usepackage{graphicx,braket,float}
\usepackage{amsmath}
\usepackage{amsfonts}
\usepackage{amsbsy}
\usepackage{xcolor}
\usepackage{soul}
\usepackage{bm}
\usepackage[normalem]{ulem}
\usepackage[unicode]{hyperref}
\hypersetup{
	unicode=true, % non-Latin characters in Acrobat?s bookmarks
	plainpages=false,
	colorlinks=true,% false: boxed links; true: colored links
	linkcolor=blue,% color of internal links
	citecolor=blue,% color of links to bibliography
	filecolor=blue,% color of file links
	urlcolor=blue% color of external links
}
\usepackage{physics}
\usepackage{bbold}
\usepackage{epstopdf}

\DeclareMathOperator{\impart}{\mathrm{Im}}

\begin{document}

%\preprint{}

\title{Stochastic Gross-Pitaevskii theory for a spin-1 Bose gas: Application to superfluidity in two dimensions}


%\author{}
%\thanks{}
%\affiliation{}

\author{Andrew P. C. Underwood}
\affiliation{Department of Physics, Centre for Quantum Science, and Dodd-Walls Centre for Photonic and Quantum Technologies, University of Otago, Dunedin, New Zealand}

\author{P. B. Blakie}
\affiliation{Department of Physics, Centre for Quantum Science, and Dodd-Walls Centre for Photonic and Quantum Technologies, University of Otago, Dunedin, New Zealand}


\date{\today}

\begin{abstract}
This paper develops and implements the stochastic projected Gross-Pitaevskii equation for spin-1 Bose gases, addressing key considerations for numerical simulations. As an application of the theory we explore equilibrium phases in a two-dimensional spin-1 gas, where quasi-long-range order emerges via a Berezinskii-Kosterlitz-Thouless transition. Our analysis includes definition of superfluid densities for both mass and spin degrees of freedom, in a manner suitable for implementation within a stochastic projected Gross-Pitaevskii equation simulation. We present a finite-temperature phase diagram for the ferromagnetic spin-1 Bose gas and identify three distinct superfluid phases: two exhibiting conventional Berezinskii-Kosterlitz-Thouless-like behavior and a novel phase that simultaneously supports independent mass and spin superflows. As temperature increases, the stability region of this novel phase shrinks. We provide characterization of the phase transitions through consideration of the spin-component densities and the unbinding of multiple types of vortices. This work provides a foundation for further studies of nonequilibrium and finite-temperature phenomena in spinor Bose gases.

\end{abstract}

%\keywords{}

\maketitle

\section{Introduction}

Over the past two decades, a collection of methods for simulating Bose gases, known as c-field methods \cite{doi:10.1080/00018730802564254}, has seen wide application, driven by experimental developments with ultra-cold gases. In these methods, the quantum field evolution can be represented using equations of motion similar in form to the Gross–Pitaevskii equation but with stochastic modifications that incorporate quantum or thermal effects. These approaches are particularly well-suited for studying finite-temperature physics. Two versions of the theory are commonly used for the finite-temperature regime: the microcanonical projected Gross-Pitaevskii equation (PGPE) \cite{PhysRevLett.87.160402,Davis2002b}, and the grand canonical stochastic projected Gross-Pitaevskii equation (SPGPE) \cite{Gardiner_2002,Gardiner_2003,PhysRevA.86.053634}. Applications of these methods include the study of critical effects and dynamics at or near the condensation transition \cite{Davis2006a,PhysRevA.81.023623,weiler_spontaneous_2008,PhysRevLett.104.160404,PhysRevA.88.063620,PhysRevA.87.063611}, equilibrium and dynamical studies of lower-dimensional Bose gases \cite{PhysRevLett.96.020404,Davis2012a,PhysRevA.86.013627,Bradley2015a,PhysRevResearch.3.013212,PhysRevA.110.053302,PhysRevResearch.6.033152}, and the dynamics of vortices and vortex lattices \cite{PhysRevA.77.033616,PhysRevLett.133.143401,PhysRevA.84.023637}.

Spinor Bose gases possess internal spin degrees of freedom, introducing a rich variety of magnetic phases and topological defects \cite{PhysRevLett.81.742,doi:10.1143/JPSJ.67.1822,stenger_spin_1998,RevModPhys.85.1191,KAWAGUCHI2012253}. Despite numerous experimental studies examining finite-temperature and nonequilibrium properties of spinor Bose gases (e.g., see \cite{stenger_spin_1998,sadler_spontaneous_2006,Vinit2013a,PhysRevLett.100.170403,PhysRevLett.110.175302,prufer_observation_2018,Weiss2019a,PhysRevLett.116.185301,prufer_condensation_2022,Huh2024a}), the application of c-field methods to these systems has so far received little attention. A notable exception is the microcanonical PGPE study presented in Ref.~\cite{Pietila2010}. Within the PGPE framework, determining the temperature is challenging \cite{Davis2003b,Davis_2005}, especially given the additional conserved quantities in the spin-1 system, making it difficult to study thermodynamic properties or to simulate temperature quenches. These difficulties are avoided in the SPGPE, where the reservoir parameters can be specified.

The formal extension of the SPGPE formalism to spinor and multi-component Bose gases was presented in Ref.~\cite{PhysRevA.90.023631}. Predating this formalism a phenomenological extension of the scalar SPGPE theory was made to simulate a spin-1 system with spin-orbit coupling \cite{PhysRevA.86.023601}, although this work did not account for the magnetic potential of the reservoir, which we discuss below. We also mention some applications of SPGPE theory to two-component Bose gases \cite{PhysRevLett.110.215302,PhysRevA.89.033631,PhysRevResearch.3.013161}.

This paper outlines the implementation of SPGPE theory for spin-1 systems and discusses issues related to their simulation and analysis. A key aspect is definition of the low-energy modes (i.e., the coherent $C$-region) which are directly simulated by the SPGPE theory. The remaining high-energy modes are treated as a reservoir and described by three intensive thermodynamic parameters: the chemical potential ($\mu$), temperature ($T$), and the magnetic potential ($\lambda$). The first two parameters appear in the SPGPE theory of scalar Bose gases, while the magnetic potential is particular to the spinor system and is associated with the conserved $z$-magnetization.

To illustrate the application of the spin-1 SPGPE, we consider the equilibrium phases of a two-dimensional (2D) spin-1 Bose gas. In general, two-dimensional systems present a challenging problem because the transition to an ordered state is prohibited in the thermodynamic limit. However, a quasi-long-range ordered state (i.e., with algebraically decaying order) that possesses rigidity (superfluidity) can arise via a topologically driven Berezinskii-Kosterlitz-Thouless (BKT) transition~\cite{Berezinskii1970,Kosterlitz_1973,PhysRevLett.40.1727}.

The interplay between spin and phase degrees of freedom in spinor Bose gases introduces additional complexity, giving rise to novel BKT-like transitions. A particularly interesting case is the broken-axisymmetric spin-1 Bose gas, occurring in a system with ferromagnetic spin-dependent interactions and subject to a small positive quadratic Zeeman energy. In recent work, we quantified aspects of the superfluid phases for this system in the easy-plane case (i.e., without $z$-magnetization) \cite{PhysRevResearch.5.L012045} and in the more general broken-axisymmetric case with nonzero $z$-magnetization \cite{PhysRevA.110.013311}. In this paper, we unify these results and present a finite-temperature phase diagram. Additionally, we develop alternative ways of characterizing the phase transitions, which may prove useful in the experimental investigation of this system: First, we show that the semi-analytic prediction relating the critical temperature and density of the BKT transition in a 2D scalar Bose gas \cite{PhysRevLett.87.270402} can be extended to accurately estimate transition temperatures in our system. Second, we analyze vortices in the spin-components and transverse spin density, and show that the unbinding transitions of these vortices coincide with the superfluid transitions. We also address how to quantify the emergence of superfluidity in c-field methods via calculations of momentum fluctuations or the long-wavelength limits of current-current correlation functions. In the spinor system, superfluidity can emerge in both mass and spin degrees of freedom, and we generalize our approach to quantify both types of superfluidity.

The structure of this paper is as follows: In Sec.~\ref{sec:formalism}, we introduce the spin-1 Bose gas, detailing the system energy functional and mean-field ground states. In Sec.~\ref{sec:sgpe}, we introduce the spin-1 SPGPE theory and discuss its implementation. We specialize the spin-1 SPGPE to a uniform 2D system in Sec.~\ref{Sec:implementation2D}, discussing relevant considerations for simulating BKT physics.
In Sec.~\ref{sec:superfluidity}, we detail the generalized superfluidity of a spin-1 Bose gas, illustrating how superfluid fractions are computed and addressing sample correlation in the calculation of expectations using time averages of SPGPE simulations. Sec.~\ref{sec:phases} presents the superfluid phases of a 2D spin-1 gas with ferromagnetic interactions at nonzero temperatures. Conclusions are provided in Sec.~\ref{sec:concs}.
 
\section{Formalism}\label{sec:formalism}

\subsection{The spin-1 Bose gas}

Within mean field theory, the spin-1 Bose gas is described by a three-component classical field~\cite{PhysRevLett.81.742,doi:10.1143/JPSJ.67.1822,PhysRevLett.80.2027,KAWAGUCHI2012253,RevModPhys.85.1191}
\begin{equation}
	\Psi(\mathbf{r}) = \left[\psi_{1}(\mathbf{r}),\psi_{0}(\mathbf{r}),\psi_{-1}(\mathbf{r})\right]^{\mathrm{T}}.
\end{equation}
Here the components $\psi_{m}$ denote amplitudes of the $m\in\left\{1,0,-1\right\}$ magnetic sub-levels. The (number) density is obtained as $n=\Psi^{\dagger}\Psi$, with the particle number $N=\int\dd{\mathbf{r}}n$ convected by the mass current
\begin{equation}
	\mathbf{J}_{n} = \frac{\hbar}{M}\impart\left(\Psi^{\dagger}\nabla\Psi\right)\label{eq:masscurrent}
\end{equation}
(here $M$ denotes the atomic mass). Similarly, the spin-density $\mathbf{F}=\left(F_{x},F_{y},F_{z}\right)$ is obtained as $\mathbf{F}=\Psi^{\dagger}\mathbf{f}\Psi$, with the magnetization $\mathbf{M}=\int\dd{\mathbf{r}}\mathbf{F}$ convected by the $\nu\in\left\{x,y,z\right\}$ spin currents
\begin{equation}
	\mathbf{J}_{\nu} = \frac{\hbar}{M}\impart\left(\Psi^{\dagger}f_{\nu}\nabla\Psi\right).\label{eq:spincurrents}
\end{equation}
Here $\mathbf{f}=\left(f_{x},f_{y},f_{z}\right)$ denotes the vector of spin-1 matrices.

The field $\Psi$ is governed by the energy functional
\begin{equation}
	E = \int\dd{\mathbf{r}}\left\{\Psi^{\dagger}\left[-\frac{\hbar^{2}\nabla^{2}}{2M}+qf_{z}^{2}\right]\Psi+\frac{g_{n}}{2}n^{2}+\frac{g_{s}}{2}|\mathbf{F}|^{2}\right\}.\label{eq:mfen}
\end{equation}
Here $q$ denotes a quadratic Zeeman shift, which acts to break degeneracy of the $m=\pm 1$ and $m=0$ sub-levels. The coupling constants $g_{n}$ and $g_{s}$ respectively quantify density and spin-density dependent interactions. The latter may be positive or negative, with $g_{s}>0$ realized in \textsuperscript{23}Na~\cite{PhysRevLett.80.2027}, and $g_{s}<0$ realized in \textsuperscript{87}Rb~\cite{PhysRevLett.87.010404} and \textsuperscript{7}Li~\cite{huh2020}. In this paper we restrict our attention to the case $g_{s}<0$; here the gas is said to exhibit ferromagnetic interactions, and the formation of local magnetization $\mathbf{F}\neq \mathbf{0}$ is energetically favored. 

Critically, the spin-density dependent interactions allow for population transfer between the three spin components, via the spin-exchange interaction where two atoms with $m=0$ interact to give a pair with $m=1$ and $m=-1$ (or vice versa). Note this process conserves $z$-magnetization $M_{z}$. It is for this reason a linear Zeeman shift need not be explicitly included in Eq.~(\ref{eq:mfen}); the conservation of $M_{z}$ nullifies the effect of an energy gap between the $m=\pm 1$ magnetic sublevels.

%Gross-Pitaevskii evolution under Eq.~(\ref{eq:mfen}) conserves particle number and $z$-magnetization, manifested by the respective invariance of $E$ under $\mathrm{U}(1)$ gauge $\Psi\to\mathrm{e}^{\mathrm{i}\theta}\Psi$ and $\mathrm{SO}(2)$ spin $\Psi\to\mathrm{e}^{-\mathrm{i}f_{z}\alpha}\Psi$ transformations.

\subsection{Ground state phases with ferromagnetic interactions}

\begin{figure}
	\includegraphics[width=\linewidth]{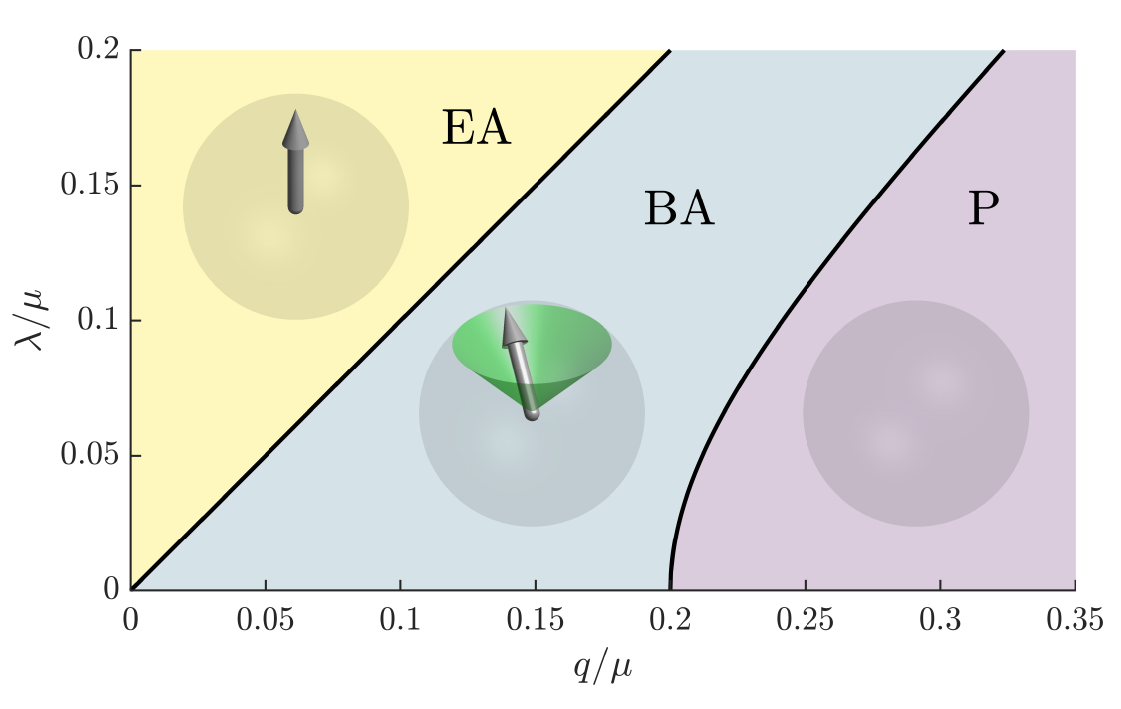}%
	\caption{Ground state phases of the spin-1 Bose gas with ferromagnetic interactions ($g_{s}=-0.1g_{n}$). Purple and yellow regions respectively mark the polar and easy-axis phases. The blue region marks the broken-axisymmetric phase, which possesses a non-zero off-axis magnetization (gray arrow).}\label{fig:groundstates}
\end{figure}
The mean-field ground states of a spin-1 Bose gas are obtained via minimization of $E-\mu N-\lambda M_{z}$, where $\mu$ and $\lambda$ respectively denote the chemical and magnetic potentials quantifying conservation of particle number $N$ and $z$-magnetization $M_{z}$. The ground state phase diagram of a gas with ferromagnetic interactions ($g_{s}=-0.1g_{n}$) is depicted in Fig.~\ref{fig:groundstates}. We discuss three distinct phases: The polar phase (P), the easy-axis phase (EA), and the broken-axisymmetric phase (BA).

\subsubsection{Polar phase (P)}
 With sufficiently large quadratic Zeeman energy, $\lambda^{2}<q^{2}-2q\mu|g_{s}/g_{n}|$, all population resides in the $m=0$ spin component. A given ground state thus has field 
 \begin{equation}
 	\Psi_{\mathrm{P}} = \sqrt{n}\mathrm{e}^{\mathrm{i}\theta}\left[0,1,0\right]^{\mathrm{T}},
 \end{equation}
and local magnetization $\mathbf{F}=\mathbf{0}$. Here the density satisfies $n = \mu/g_{n}$. The angle $\theta$ arises from the breaking of $\mathrm{U}(1)$ gauge symmetry.

\subsubsection{Easy-axis phase (EA)}
Within the easy-axis phase the magnetic potential dominates, $\lambda>q$, and all population resides in the $m=1$ spin component. A given ground state then has field
\begin{equation}
	\Psi_{\mathrm{EA}} = \sqrt{n}\mathrm{e}^{\mathrm{i}\phi}\left[1,0,0\right]^{\mathrm{T}},
\end{equation}
and local magnetization $\mathbf{F}=n\hat{\mathbf{z}}$. Here the density satisfies $n = \left(\mu+|\lambda|-q\right)/\left(g_{n}+g_{s}\right)$. This ground state is modified by both gauge transformations $\Psi_{\mathrm{EA}}\to\mathrm{e}^{\mathrm{i}\theta}\Psi_{\mathrm{EA}}$ and $z$-spin rotations $\Psi_{\mathrm{EA}}\to\mathrm{e}^{-\mathrm{i}f_{z}\alpha}\Psi_{\mathrm{EA}}$. However, with the gas axially magnetized these act identically, and only a single $\mathrm{U}(1)$ symmetry is broken, made explicit by the angle $\phi=\theta-\alpha$.

\subsubsection{Broken-axisymmetric phase (BA)}\label{Sec:BA}
Within the broken-axisymmetric phase all spin components are occupied, and a given ground state has the form 
\begin{equation}
	\Psi_{\mathrm{BA}} = \sqrt{n}\mathrm{e}^{\mathrm{i}\theta}\mathrm{e}^{-\mathrm{i}f_{z}\alpha}\xi,
\end{equation}
where $\xi$ is the normalised spinor~\cite{PhysRevA.75.013607,KAWAGUCHI2012253}
\begin{equation}
	\xi = \begin{bmatrix}
		\dfrac{q+\lambda}{2q}\sqrt{\dfrac{\lambda^{2}-q^{2}+2|g_{s}|nq}{2|g_{s}|nq}}\\
		\sqrt{\dfrac{\left(q^{2}-\lambda^{2}\right)\left(\lambda^{2}+q^{2}+2|g_{s}|nq\right)}{4|g_{s}|nq^{3}}}\\
		\dfrac{q-\lambda}{2q}\sqrt{\dfrac{\lambda^{2}-q^{2}+2|g_{s}|nq}{2|g_{s}|nq}}
	\end{bmatrix}.
\end{equation}
The density satisfies $n = \left(\mu-q/2+\lambda^{2}/2q\right)/\left(g_{n}+g_{s}\right)$. Reflected by the parameters $\theta$ and $\alpha$, such a ground state breaks both $\mathrm{U}(1)$ gauge and $\mathrm{SO}(2)$ spin-rotational symmetries. The latter arises from the realization of an arbitrary orientation of the transverse magnetization
\begin{equation}
	\mathbf{F}_{\perp}=\left(F_{x},F_{y}\right) = |\mathbf{F}_{\perp}|\left(\cos\alpha,\sin\alpha\right).\label{eq:Fperp}
\end{equation}
With $\lambda=0$ the BA ground state exhibits planar magnetization, $F_{z}=0$. As $\lambda$ increases $\mathbf{F}$ tilts out of the plane, with
\begin{equation}
	F_{z} = \frac{\lambda\left(\lambda^{2}-q^{2}+2|g_{s}|nq\right)}{2|g_{s}|q^{2}},
\end{equation}
until the gas is fully magnetized at the BA - EA phase boundary, $\lambda=q$.

\subsection{Non-zero temperature}\label{Sec:NZT}
At non-zero temperatures, within a grand-canonical treatment, the spin-1 Bose gas is governed by partition function
\begin{equation}
	Z = \int\mathrm{D}\Psi\,\mathrm{e}^{-\left(E[\Psi]-\mu N[\Psi]-\lambda M_{z}[\Psi]\right)/k_{B}T},\label{eq:Z}
\end{equation}
where $E[\Psi]$ denotes the energy (\ref{eq:mfen}), and the integration is over classical fields $\Psi(\mathbf{r})$, which now constitute microstates. Equilibrium expectation values of an observable $\mathcal{O}[\Psi]$ are given by
\begin{equation}
	\langle\mathcal{O}\rangle = \frac{1}{Z}\int\mathrm{D}\Psi\,\mathcal{O}[\Psi]\mathrm{e}^{-\left(E[\Psi]-\mu N[\Psi]-\lambda M_{z}[\Psi]\right)/k_{B}T}.\label{eq:expval}
\end{equation}
The evaluation of such expectation values is a primary goal of the SPGPE model, which we discuss in the following section.

\section{Stochastic Gross-Pitaevskii model of a spin-1 Bose gas}\label{sec:sgpe}

\subsection{Construction of the classical field}

We investigate the finite-temperature properties of the spin-1 Bose gas via a c-field model~\cite{doi:10.1080/00018730802564254}. Such a treatment is predicated on the notion that within a degenerate Bose gas near equilibrium, many low-energy modes will have appreciable occupation (i.e. of at least order unity), permitting their description by a classical field. Formally, we take the eigenstates $\varphi_{\mathbf{n},m}(\mathbf{r})$ of the single-particle Hamiltonian [cf. Eq.~(\ref{eq:mfen})]
\begin{equation}
	H_{\mathrm{sp}} = -\frac{\hbar^{2}\nabla^{2}}{2M}+V_{m}(\mathbf{r})+qm^{2}
\end{equation}
as a basis, and denote by $\hat{a}_{\mathbf{n},m}$ the bosonic operator annihilating a particle with magnetic quantum number $m$ in spatial mode $\mathbf{n}$ [here $V_{m}(\mathbf{r})$ denotes a component-dependent external potential]. Modes $\mathbf{n}$ satisfying $\langle\hat{a}^{\dagger}_{\mathbf{n},m}\hat{a}_{\mathbf{n},m}\rangle\gtrsim 1$ are suitably treated with the prescription $\hat{a}_{\mathbf{n},m}\to c_{\mathbf{n},m}$, where $c_{\mathbf{n},m}$ is a complex amplitude~\cite{PhysRevLett.87.160402,PhysRevA.66.051602}. Collectively then, these modes are described by the classical field
\begin{equation}
	\psi_{m}(\mathbf{r}) = \sum_{\mathbf{n}\in C}c_{\mathbf{n},m}\varphi_{\mathbf{n},m}(\mathbf{r}).
\end{equation}
Here the restriction of summation to the so-called $C$-region, defined by $C = \left\{\mathbf{n}\colon\langle|c_{\mathbf{n},m}|^{2}\rangle\gtrsim 1\right\}$, makes explicit the condition of sufficient mode occupation.

The remaining sparingly occupied modes $\mathbf{n}\notin C$ constitute the incoherent region ($I$-region). Generally, such modes are weakly interacting, and exhibit fast dynamics compared to the $C$-region, so that they may be well-approximated as a thermalized reservoir. The effect of the $I$-region on the classical field dynamics manifests as noise and damping terms in the $C$-field evolution, as we detail in the next subsection.

\subsection{Time evolution}\label{sec:timeev}

%\textcolor{red}{The time evolution of $\Psi(\mathbf{r})$ is described by a spin-1 SPGPE equation. The theory is developed   in Ref.~\cite{PhysRevA.90.023631} where the system Hamiltonian is subdivided into 
	%$C$-region  and $I$-region parts and a Master equation is developed for the $C$-region evolution. Using the Wigner function to represent the density matrix and approximate its evolution as a Fokker-Planck equation, the SPGPE emerges as the mapping to an equivalent stochastic differential equation. 
	%Within the simple growth approximation the  spin-1 SPGPE equation takes the form}

The time evolution of $\Psi(\mathbf{r})$ is described by a spin-1 SPGPE equation, the theory of which is developed in Ref.~\cite{PhysRevA.90.023631}: here the system Hamiltonian is subdivided into 
$C$-region and $I$-region parts, with a Master equation developed for the $C$-region evolution. Representing the $C$-region density matrix with a Wigner function, the approximate $C$-region evolution is then expressed with a Fokker-Planck equation, from which the spin-1 SPGPE emerges as the mapping to an equivalent stochastic differential equation. Within the simple growth approximation this takes the form
\begin{equation}
	\mathrm{i}\hbar\mathrm{d}\Psi = \mathcal{P}\left[\left(1-\mathrm{i}\gamma\right)\left(\mathcal{L}-\mu-\lambda f_{z}\right)\Psi\mathrm{d}t+\mathrm{i}\hbar\mathrm{d}W_{\gamma}\right].\label{eq:sgpe}
\end{equation}
The projector $\mathcal{P}$ enforces that all modes maintain macroscopic occupation throughout time evolution; its action on an arbitrary field $f(\mathbf{r})=\sum_{m}f_{m}(\mathbf{r})\xi_{m}$ is
\begin{equation}
	\mathcal{P}[f(\mathbf{r})] = \sum_{m}\xi_{m}\int\dd{\mathbf{r}'}\delta_{m}(\mathbf{r},\mathbf{r}')f_{m}(\mathbf{r}'),
\end{equation}
where
\begin{equation}
	\delta_{m}(\mathbf{r},\mathbf{r}') = \sum_{\mathbf{n}\in C}\varphi_{\mathbf{n},m}(\mathbf{r})\varphi_{\mathbf{n},m}^{*}(\mathbf{r}')
\end{equation}
is an incomplete delta function. Here $\xi_{m}$ denote the $f_{z}$ spin matrix eigenstates [$\left(\xi_{m}\right)_{m'}=\delta_{m,m'}$]. The nonlinear operator $\mathcal{L}$ describes Gross-Pitaevskii evolution, acting as
\begin{equation}
	\mathcal{L}\Psi = \left(-\frac{\hbar^{2}\nabla^{2}}{2M}+qf_{z}^{2}+g_{n}n+g_{s}\sum_{\nu}F_{\nu}f_{\nu}\right)\Psi.
\end{equation}
Terms containing the dimensionless parameter $\gamma>0$ arise from interactions with the high-energy ($I$-region) atoms not described by field $\Psi(\mathbf{r})$. In particular, they model the growth process wherein two $I$-region atoms interact with the effect of adding one atom to the $C$-region, allowing for exchange of energy, particle number and $z$-magnetization.
In general the reservoir coupling should be described by a number of distinct parameters reflecting the different spin-channels, which can be obtained analytically under some approximation \cite{PhysRevA.90.023631}. We note that in practical applications to scalar Bose-gas experiments, these couplings are often informed by direct comparison to experiment (e.g.~by matching condensate growth rates in cooling \cite{weiler_spontaneous_2008}). However, the equilibrium properties of interest here are independent of these couplings, permitting our simplified description with a single parameter $\gamma$.

 The components of $\mathrm{d}W_{\gamma} = \left[\mathrm{d}W_{1},\mathrm{d}W_{0},\mathrm{d}W_{-1}\right]^{\mathrm{T}}$ are circularly-symmetric complex Gaussian noise with correlations
\begin{equation}
	\langle\mathrm{d}W_{m}^{*}(\mathbf{r})\mathrm{d}W_{m'}(\mathbf{r}')\rangle = \frac{2\gamma k_{B}T}{\hbar}\delta_{m}(\mathbf{r},\mathbf{r}')\delta_{m,m'}\mathrm{d}t.
\end{equation}
Independent of the value of $\gamma$, steady-state solutions of Eq.~(\ref{eq:sgpe}) sample a grand-canonical reservoir with temperature $T$, chemical potential $\mu$, and magnetic potential $\lambda$. Consequently, equilibrium expectation values (\ref{eq:expval}) may be approximated via time averages: Computing samples $\mathcal{O}_{i}=\mathcal{O}[\Psi(t_{i})]$ of the observable $\mathcal{O}$ at times $t_{i}$, $i\in\left\{1,2,\ldots,\mathcal{N}_{s}\right\}$, we have $\langle\mathcal{O}\rangle\approx\bar{\mathcal{O}}$ with
\begin{equation}
	\bar{\mathcal{O}}= \frac{1}{\mathcal{N}_{s}}\sum_{i=1}^{\mathcal{N}_{s}}\mathcal{O}_{i}.\label{eq:timeav}
\end{equation}
This procedure is reliant on an appropriate method of sampling; this will be discussed further in Sec.~\ref{Sec:SFcalc}.

\section{Implementation for a uniform 2D spin-1 Bose gas}\label{Sec:implementation2D}
Here we apply the spin-1 SPGPE to describe the equilibrium properties of a 2D system. We hereon restrict $\mathbf{r}\to(x,y)$, set $V_{m}(\mathbf{r})=0$ \footnote{In practice a quasi-2D gas can be produced in experiments by the application of a tight harmonic trapping potential, i.e.~$V_m(\mathbf{r}) = M\omega^{2}z^{2}/2$. The effective 2D coupling constants $g_{n}$ and $g_{s}$ differ from their 3D values by a factor of $\sqrt{M\omega/2\pi\hbar}$~\cite{PhysRevLett.99.040402,PhysRevLett.102.170401}.}, and simulate the system on a square domain of dimensions $L\times L$, with periodic boundary conditions. Doing so, the single particle modes are plane-waves
\begin{align}
\varphi_{\mathbf{n},m}=\left(1/L\right)\mathrm{e}^{\mathrm{i}\mathbf{k}_{\mathbf{n}}\cdot\mathbf{r}},
\end{align}
 with $\mathbf{k}_{\mathbf{n}}=\left(2\pi/L\right)\mathbf{n}$, and $\mathbf{n}\in\mathbb{Z}\times\mathbb{Z}$. The three-component field thus takes the form
\begin{equation}
	\Psi(\mathbf{r}) = \frac{1}{L}\sum_{\mathbf{n}\in C}\sum_{m}c_{\mathbf{n},m}\xi_{m}\mathrm{e}^{\mathrm{i}\mathbf{k}_{\mathbf{n}}\cdot\mathbf{r}}.\label{eq:cfield}
\end{equation}
This construction has the advantage that $\Psi(\mathbf{r})$ can be represented efficiently on a mesh of equally spaced points, with the SPGPE evolution (\ref{eq:sgpe}) efficiently implemented using Fast Fourier transformations (e.g.~see \cite{Davis2002b,Blakie2008b}).

\subsubsection{Choice of C-region}\label{Sec:Cregion}
Implementation of this model requires an appropriate choice of $C$-region. All strongly interacting modes must be included in the field $\Psi(\mathbf{r})$, so as to provide an adequate description of their nonperturbative dynamics. Simultaneously, the field $\Psi(\mathbf{r})$ must contain no modes with sub-unity occupation, so as to justify the classical field treatment. Comparing kinetic and interaction energies in Eq.~(\ref{eq:mfen}) one finds strongly and weakly interacting modes are demarcated by wavenumber~\cite{PhysRevLett.87.270402,PhysRevA.66.043608}
\begin{equation}
	k_{\text{int}} = \frac{\sqrt{2M\mu}}{\hbar}.
\end{equation}
Modes $\mathbf{n}$ with wavenumbers $|\mathbf{k}_{\mathbf{n}}|<k_{\text{int}}$ are strongly interacting, and must be included in $\Psi(\mathbf{r})$. Contrarily, modes with $|\mathbf{k}_{\mathbf{n}}|\gtrsim k_{\text{int}}$ are effectively non-interacting. With energies $E_{\mathbf{n}}\approx \sum_{m}|c_{\mathbf{n},m}|^{2}\left(\hbar^{2}|\mathbf{k}_{\mathbf{n}}|^{2}/2M\right)$, the non-interacting modes have equilibrium occupations $N_{\mathbf{n},m}=\langle|c_{\mathbf{n},m}|^{2}\rangle$ given by
\begin{equation}
	N_{\mathbf{n},m}\approx \frac{2Mk_{B}T}{\hbar^{2}|\mathbf{k}_{\mathbf{n}}|^{2}}.\label{eq:equidist}
\end{equation}
This yields a prescription for the choice of $C$-region: The field $\Psi(\mathbf{r})$ may contain all modes with
\begin{equation}
	|\mathbf{k}_{\mathbf{n}}|\lesssim k_{\text{cut}} = \frac{\sqrt{2Mk_{B}T}}{\hbar},
\end{equation}
and must contain all those with $|\mathbf{k}_{\mathbf{n}}|<k_{\text{int}}$. Note that these conditions may be simultaneously satisfied only if $k_{B}T\gtrsim\mu$.

We implement this prescription through our choice of numerical grid. In particular, we simulate $\mathcal{N}=\mathcal{N}_{x}\times\mathcal{N}_{x}$ modes $\mathbf{n}=\left(n_{x},n_{y}\right)$ labelled by indices $n_{x},n_{y}\in\left\{-\mathcal{N}_{x}/2,-\mathcal{N}_{x}/2+1,\ldots,\mathcal{N}_{x}/2-1\right\}$, with temperature-dependent system size
\begin{equation}
	L = \mathcal{N}_{x}\sqrt{\frac{2\pi\hbar^{2}}{Mk_{B}T}}.
\end{equation}
This treatment has the virtue of reducing the projector $\mathcal{P}$ to the identity operator, simplifying evolution of Eq.~(\ref{eq:sgpe}).

\subsubsection{Noise generation}

The noise components $\mathrm{d}W_{m}(\mathbf{r})$ may be generated by noting that the amplitudes
\begin{equation}
	\mathrm{d}W_{\mathbf{n},m} =\frac{1}{L}\int\dd{\mathbf{r}}\mathrm{e}^{-\mathrm{i}\mathbf{k}_{\mathbf{n}}\cdot\mathbf{r}}\mathrm{d}W_{m}(\mathbf{r})
\end{equation}
have correlations
\begin{equation}
	\langle \mathrm{d}W_{\mathbf{n},m}^{*}\mathrm{d}W_{\mathbf{m},m'}\rangle = \frac{2\gamma k_{B}T}{\hbar}\delta_{\mathbf{n},\mathbf{m}}\delta_{m,m'}\mathrm{d}t.
\end{equation}
That is, each component $m$ of each mode $\mathbf{n}$ has independent noise sampled from a circularly-symmetric complex Gaussian distribution with standard deviation
\begin{equation}
	\sigma = \sqrt{\frac{2\gamma k_{B}T}{\hbar}}\sqrt{\mathrm{d}t}.
\end{equation}

\subsubsection{Choice of units}

We take the chemical potential $\mu$ as an energy unit, and introduce the corresponding length and time units, respectively $x_{\mu} =\hbar/\sqrt{M\mu}$ and $t_{\mu}=\hbar/\mu$. The interaction strengths are quantified by the dimensionless parameters $\tilde{g}_{n}=Mg_{n}/\hbar^{2}$ and $\tilde{g}_{s}=Mg_{s}/\hbar^{2}$; we take $\tilde{g}_{n}=0.15$, motivated by Refs.~\cite{PhysRevLett.99.040402,PhysRevLett.121.145301}, and $\tilde{g}_{s}=-0.1\tilde{g}_{n}$. We set $\gamma =0.1$; this value is much larger than that expected from a formal calculation (see Ref.~\cite{doi:10.1080/00018730802564254}), chosen so as to hasten field thermalization, while being sufficiently small so that the noise $\mathrm{d}W_{\gamma}$ may be numerically treated with an Euler step~\cite{milstein_stochastic_2003}. In the following we compute the dependence of the system superfluid properties on the scaled temperature
\begin{equation}
	\mathcal{T}=\frac{Mg_{n}k_{B}T}{\hbar^{2}\mu}.
\end{equation}

\subsubsection{Equilibration}

\begin{figure}
	\includegraphics[width=\linewidth]{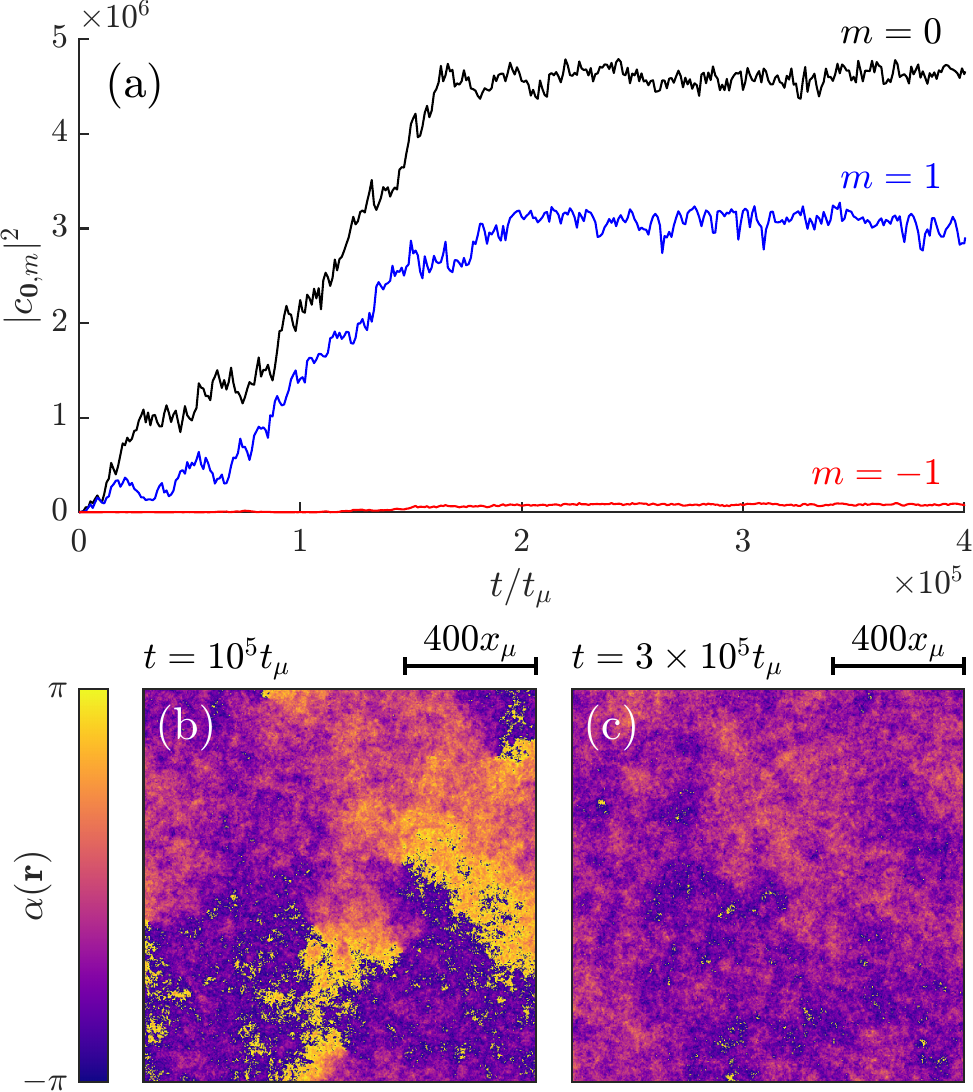}
	\caption{Attainment of a steady state with the stochastic Gross-Pitaevskii model. (a) Instantaneous occupation of the $\mathbf{n}=\mathbf{0}$ mode during a single trajectory of the spin-1 SPGPE, beginning from an empty initial state. (b) and (c): Spatial profiles of the spin angle $\alpha(\mathbf{r})$ after evolution times of $10^{5}t_{\mu}$ and $3\times 10^{5}t_{\mu}$, respectively. Both panels have dimensions of $1200x_{\mu}\times 1200x_{\mu}$. Results obtained with parameters $q=0.1\mu$, $\lambda=0.06\mu$, $\mathcal{T}=0.15$, and $\mathcal{N}_{x}=512$.}\label{fig:therm}
\end{figure}

We obtain steady states by evolving an initially empty field $\Psi(\mathbf{r}) = \left[0,0,0\right]^{\mathrm{T}}$ until saturation of the $\mathbf{k}=\mathbf{0}$ mode occupation is observed, at which point we consider the field to be thermalized. The growth of this mode during a typical thermalization procedure is depicted in Fig.~\ref{fig:therm}.~(a). In general, the thermalization time increases with increasing system size and decreasing temperature. Here, with $\mathcal{T}=0.15$ and $\mathcal{N}_{x}=512$, we observe saturation of the instantaneous occupation $|c_{\mathbf{0},m}|^{2}$ after an evolution time of $2\times 10^{5}t_{\mu}$. Due to the choice of empty initial state, immediately following the initialization of evolution the field $\Psi(\mathbf{r})$ is disordered, being comprised only of noise $\mathrm{d}W_{\gamma}$. As such, the transient dynamics are dominated by the annihilation of vortex-antivortex pairs, and the subsequent decay of vortex density towards its equilibrium value. In Figs.~\ref{fig:therm}.~(b) and (c) we show the spin angle $\alpha(\mathbf{r})$ (\ref{eq:Fperp}) after evolution times of $10^{5}t_{\mu}$ and $3\times 10^{5}t_{\mu}$, respectively. Multiple vortices (and antivortices) are observed during thermalization, identified as points around which the spin angle winds by $\pm2\pi$. In this case, these vortices are not thermally excited, but rather are remnants of the disordered initial state. Following thermalization these vortices have annihilated.

\begin{figure}
	\includegraphics[width=\linewidth]{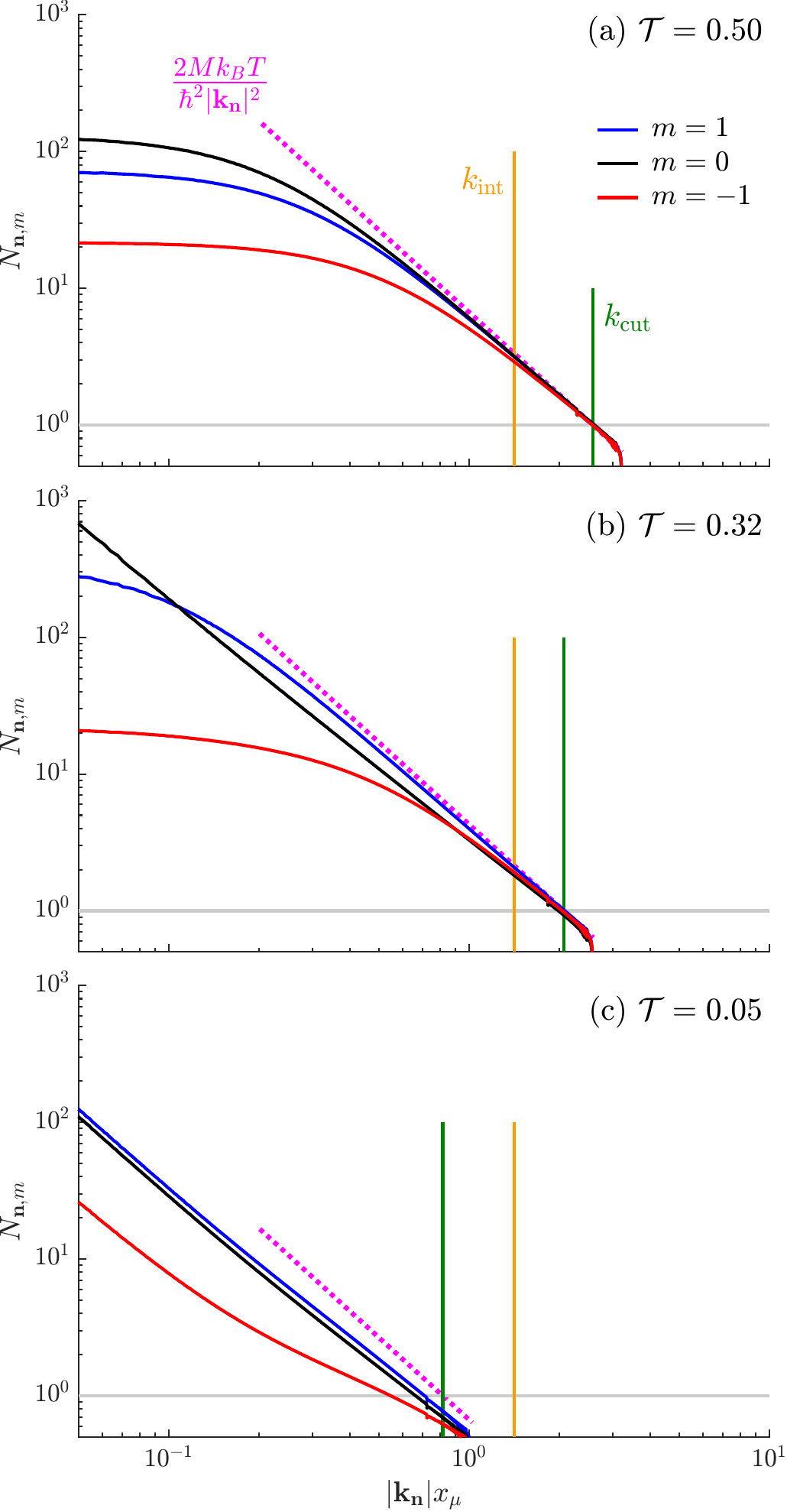}
	\caption{Equilibrium mode occupations $N_{\mathbf{n},m}$ as a function of wavenumber $|\mathbf{k}_{\mathbf{n}}|$, at (a) $\mathcal{T}=0.50$, (b) $\mathcal{T}=0.32$, and (c) $\mathcal{T}=0.05$. Blue, black, and red lines respectively denote the occupations of the $m=1$, $m=0$, and $m=-1$ spin components. Magenta dotted lines denote the non-interacting result (\ref{eq:equidist}). Vertical green and orange lines respectively denote $k_{\text{cut}}$ and $k_{\text{int}}$. All results obtained with $q=0.1\mu$ and $\lambda=0.06\mu$.}\label{fig:modepops}
\end{figure}

Equilibrium mode occupations $N_{\mathbf{n},m}$ are depicted in Fig.~\ref{fig:modepops}. At high temperatures $\mathcal{T}\gtrsim \tilde{g}_{n}$ (i.e. $k_{B}T\gtrsim\mu$) we observe the expected behavior (\ref{eq:equidist}) at $|\mathbf{k}_{\mathbf{n}}|>k_{\text{int}}$. This comparison is a useful test for validating that the noise and damping terms have been implemented correctly in the SPGPE theory. At lower temperatures $\mathcal{T}<\tilde{g}_{n}$ interactions remain significant at the high-wavenumber cutoff, and validity of this classical field treatment breaks down.

\section{Spin-1 superfluidity}\label{sec:superfluidity}

\subsection{Superfluid measures}

Spin-1 Bose gases support superflows of both mass (\ref{eq:masscurrent}) and spin currents (\ref{eq:spincurrents}). Such flows may be investigated by considering the system rigidity against spatial variations of its symmetry-broken angles. Motivated by the ground state phases previously discussed, we allow for variation in the global phase $\theta(\mathbf{r})$ and transverse spin angle $\alpha(\mathbf{r})$. This will increase the system free energy by an amount $\Delta F$ with quadratic form
\begin{equation}
	\Delta F = \frac{\hbar^{2}}{2M}\int\dd{\mathbf{r}}\left[\rho_{nn}|\nabla\theta|^{2}+\rho_{ss}|\nabla\alpha|^{2}+2\rho_{ns}\nabla\theta\cdot\nabla\alpha\right].\label{eq:delF}
\end{equation}
The coefficients $\rho_{nn}$ and $\rho_{ss}$ respectively define the mass ($n$) and spin ($s$) superfluid densities. The coefficient $\rho_{ns}$ simultaneously describes both mass and spin superflows; it is bound by the condition that $\Delta F$ be strictly positive, satisfying $\rho_{ns}\leq\sqrt{\rho_{nn}\rho_{ss}}$.

We engineer spatial variations in the symmetry-broken phases $\theta$ and $\alpha$ by transformation of the fields $\Psi(\mathbf{r})$. In particular, taking
\begin{equation}
	\Psi(\mathbf{r})\to\mathrm{e}^{\mathrm{i}\kappa_{n}\hat{\mathbf{n}}\cdot\mathbf{r}}\mathrm{e}^{\mathrm{i}f_{z}\kappa_{s}\hat{\mathbf{n}}\cdot\mathbf{r}}\Psi(\mathbf{r}),\label{eq:transform}
\end{equation}
with $\hat{\mathbf{n}}$ an arbitrary unit vector, one obtains the equilibrium phase gradients $\nabla\theta = \kappa_{n}\hat{\mathbf{n}}$ and $\nabla\alpha=\kappa_{s}\hat{\mathbf{n}}$. At zero temperature the resultant change in free energy is given by Eq.~(\ref{eq:mfen}), so that
\begin{align}
	\rho_{nn} &= n,\\
	\rho_{ss} &= \Psi^{\dagger}f_{z}^{2}\Psi,\\
	\rho_{ns} &= F_{z}.
\end{align}
At non-zero temperature, the resultant change in free energy $F=-k_{B}T\ln Z$ is obtained from the partition function (\ref{eq:Z}). We again write $Z$ as an integral over un-twisted fields $\Psi(\mathbf{r})$; the phase twists present in (\ref{eq:transform}) then manifest in the transformation
\begin{equation}
	E\to E +\sum_{i,j}\frac{\hbar^{2}\kappa_{i}\kappa_{j}}{2M}N_{ij}+\sum_{i}\frac{\hbar\kappa_{i}}{M}\mathbf{P}_{i}\cdot\hat{\mathbf{n}},\label{eq:Etrans}
\end{equation}
making explicit the increase in kinetic energy arising from the equilibrium phase gradients. Here summation is performed over $i,j\in\left\{n,s\right\}$. Quantities $N_{nn}$ and $N_{ns}$ are respectively the particle number $N$ and $z$-magnetization $M_{z}$, while $N_{ss}$ is the sum occupation of the $m=\pm 1$ spin components, i.e. $N_{ss}=\int\dd{\mathbf{r}}\Psi^{\dagger}f_{z}^{2}\Psi$. The momenta are given by $\mathbf{P}_{i}=M\int\dd{\mathbf{r}}\mathbf{J}_{i}(\mathbf{r})$, with $\mathbf{J}_{s}$ denoting the $z$-spin current (\ref{eq:spincurrents}). Equation (\ref{eq:Etrans}) identifies the $\kappa_{i}$ dependence of the free energy. With this, the superfluid densities of a 2D fluid confined to area $L\times L$ may be evaluated as [see Eq.~(\ref{eq:delF})]
\begin{equation}
	\rho_{ij} = \frac{M}{\hbar^{2}L^{2}}\frac{\partial^{2}F}{\partial\kappa_{i}\partial\kappa_{j}}\bigg|_{\kappa_{i}=\kappa_{j}=0}.
\end{equation}
This gives the result
\begin{equation}
	\rho_{ij} = \frac{\langle N_{ij}\rangle}{L^{2}} - \frac{\hat{\mathbf{n}}^{\mathrm{T}}\langle\mathbf{P}_{i}\mathbf{P}_{j}^{\mathrm{T}}\rangle\hat{\mathbf{n}}}{Mk_{B}TL^{2}},\label{eq:rhoij}
\end{equation}
where the expectation values are evaluated in a system with periodic boundary conditions (i.e. with no enforced phase twists). The first term of Eq.~(\ref{eq:rhoij}) is interpreted as a generalized total density, $n_{ij} = \langle N_{ij}\rangle/L^{2}$. The second is interpreted as a non-superfluid `normal' density. Averaging over the arbitrary direction $\hat{\mathbf{n}}$, this reads
\begin{equation}
	\varrho_{ij} = \frac{\langle\mathbf{P}_{i}\cdot\mathbf{P}_{j}\rangle}{2Mk_{B}TL^{2}}.\label{eq:varrho}
\end{equation}

Implicit in the second term of Eq.~(\ref{eq:rhoij}) are the integrated current correlations
\begin{equation}
	\varrho_{ij} = \frac{M}{k_{B}TL^{2}}\int\dd{\mathbf{r}}\int\dd{\mathbf{r}'}\hat{\mathbf{n}}^{\mathrm{T}}\langle\mathbf{J}_{i}(\mathbf{r})\mathbf{J}_{j}^{\mathrm{T}}(\mathbf{r}')\rangle\hat{\mathbf{n}}.
\end{equation}
Within a sufficiently large system the integrand may be assumed isotropic, being dependent only on the separation $\mathbf{r}-\mathbf{r}'$. In this case, one may write
\begin{equation}
	\varrho_{ij} = \int\dd{\mathbf{k}}\Delta(\mathbf{k})\hat{\mathbf{n}}^{\mathrm{T}}\chi_{ij}(\mathbf{k})\hat{\mathbf{n}},\label{eq:rhoijJJ}
\end{equation}
where $\Delta(\mathbf{k}) = \left(1/4\pi^{2}\right)\int\dd{\mathbf{r}}\exp(\mathrm{i}\mathbf{k}\cdot\mathbf{r})$, and $\chi_{ij}(\mathbf{k})$ are the response tensors
\begin{equation}
	\chi_{ij}(\mathbf{k}) = \frac{M}{k_{B}T}\int\dd{\mathbf{r}}\mathrm{e}^{-\mathrm{i}\mathbf{k}\cdot\mathbf{r}}\langle\mathbf{J}_{i}(\mathbf{0})\mathbf{J}_{j}^{\mathrm{T}}(\mathbf{r})\rangle.
\end{equation}
Taking the infinite system size limit, $\Delta(\mathbf{k})$ approaches a delta-function $\delta(\mathbf{k})$, so that $\varrho_{ij}$ is determined from a long-wavelength limit of the response tensor $\chi_{ij}(\mathbf{k})$. However, the limiting procedure must be chosen appropriately: Suppose the system is initially confined to a rectangular box with walls of length $L_{n}$ and $L_{m}$, aligned along directions $\hat{\mathbf{n}}$ and $\hat{\mathbf{m}}\perp\hat{\mathbf{n}}$ respectively. Writing $\mathbf{k}=k_{n}\hat{\mathbf{n}}+k_{m}\hat{\mathbf{m}}$, the function $\Delta(\mathbf{k})$ may be separated as $\Delta(\mathbf{k})=\Delta_{n}(k_{n})\Delta_{m}(k_{m})$, where $\Delta_{n}(k)\to \delta(k)$ as $L_{n}\to\infty$, and $\Delta_{m}(k)\to \delta(k)$ as $L_{m}\to\infty$. Suppose we first let $L_{n}\to\infty$ before subsequently letting $L_{m}\to\infty$. Applying the above separation of $\Delta(\mathbf{k})$, we obtain
\begin{equation}
\int\dd{\mathbf{k}}\Delta(\mathbf{k})\hat{\mathbf{n}}^{\mathrm{T}}\chi_{ij}(\mathbf{k})\hat{\mathbf{n}}\to\lim_{k\to 0}\hat{\mathbf{n}}^{\mathrm{T}}\chi_{ij}(k\hat{\mathbf{m}})\hat{\mathbf{n}}.\label{eq:lim1}
\end{equation}
Alternatively, letting $L_{m}\to\infty$ before subsequently letting $L_{n}\to\infty$ we obtain
\begin{equation}
	\int\dd{\mathbf{k}}\Delta(\mathbf{k})\hat{\mathbf{n}}^{\mathrm{T}}\chi_{ij}(\mathbf{k})\hat{\mathbf{n}}\to\lim_{k\to 0}\hat{\mathbf{n}}^{\mathrm{T}}\chi_{ij}(k\hat{\mathbf{n}})\hat{\mathbf{n}}.\label{eq:lim2}
\end{equation}
The results (\ref{eq:lim1}) and (\ref{eq:lim2}) are in general not equivalent; the discrepancy between them is a key property of a superfluid. By analogy with the theory of single-component superfluids~\cite{PhysRevA.81.023623,Baym1968}, we identify Eq.~(\ref{eq:lim1}) as the appropriate result for evaluation of the normal densities $\varrho_{ij}$. Also by analogy, we surmise that the result (\ref{eq:lim2}) gives the generalized total densities $n_{ij}$. 

Being dependent only on $\mathbf{k}$, the tensors $\chi_{ij}(\mathbf{k})$ are constructed as a linear combination of the $2\times 2$ identity matrix $\mathbb{1}$, and $\mathbf{k}\mathbf{k}^{\mathrm{T}}$. As such, they are fully characterized by their longitudinal $\chi_{ij}^{L}(k)$ and transverse $\chi_{ij}^{T}(k)$ components as~\cite{Baym1968}
\begin{equation}
	\chi_{ij}(\mathbf{k}) =\frac{\mathbf{k}\mathbf{k}^{\mathrm{T}}}{k^{2}}\chi_{ij}^{L}(k)+\left(\mathbb{1}-\frac{\mathbf{k}\mathbf{k}^{\mathrm{T}}}{k^{2}}\right)\chi_{ij}^{T}(k)
\end{equation}
(here $k=|\mathbf{k}|$). With this, the results (\ref{eq:lim1}) and (\ref{eq:lim2}) may be expressed as
\begin{align}
	\varrho_{ij} &= \lim_{k\to 0}\chi_{ij}^{T}(k),\\
	n_{ij} &= \lim_{k\to 0}\chi_{ij}^{L}(k),
\end{align}
giving the alternative measure of superfluidity
\begin{equation}
	\rho_{ij} = \lim_{k\to 0}\left[\chi_{ij}^{L}(k)-\chi_{ij}^{T}(k)\right].\label{eq:sffromchi}
\end{equation}

\subsection{Superfluid density calculations}\label{Sec:SFcalc}

\begin{figure}
	\includegraphics[width=\linewidth]{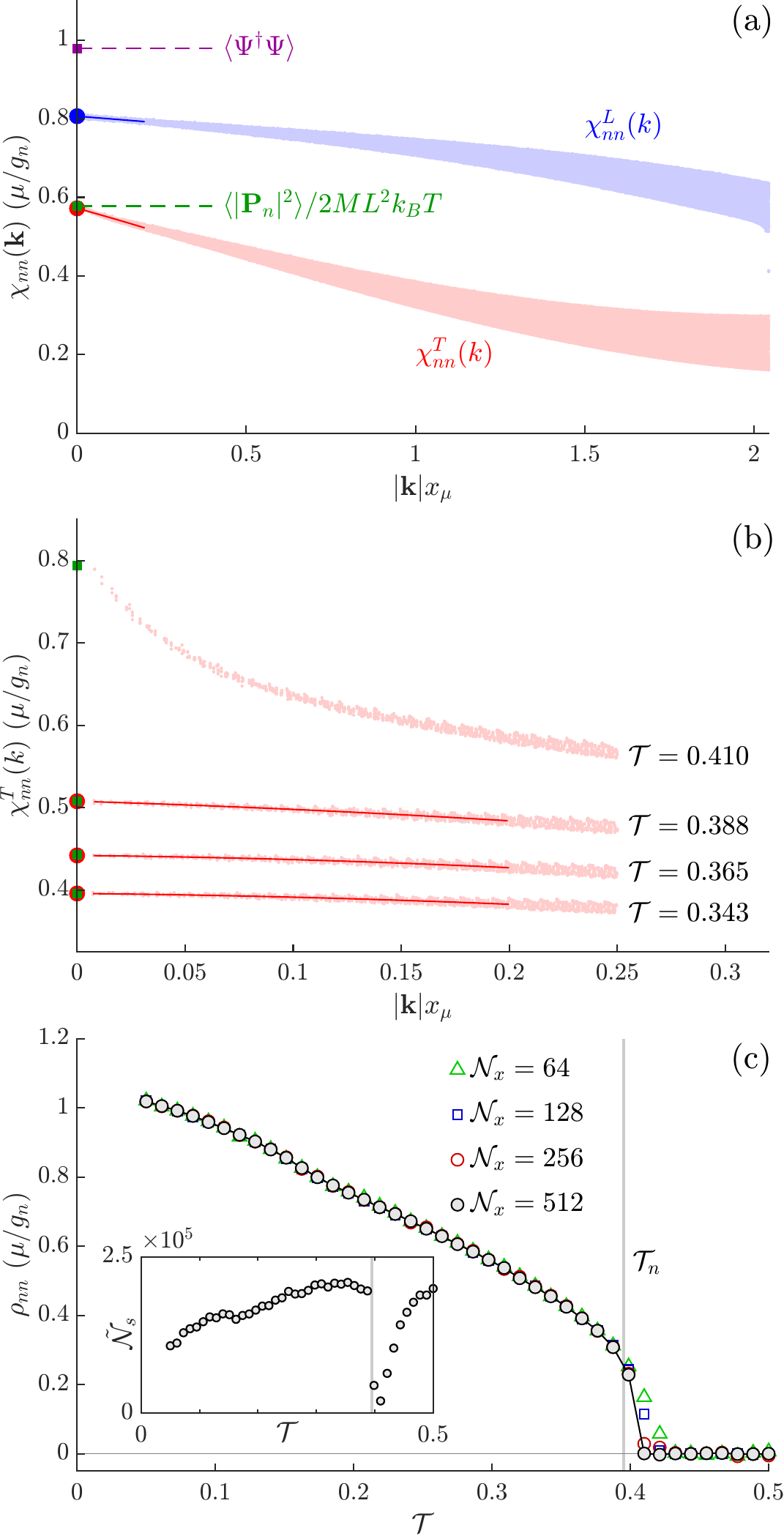}
	\caption{Computation of superfluid densities. All results obtained with $q=0.1\mu$ and $\lambda=0$. (a) Extraction of normal $\varrho_{nn}$ and total $n_{nn}$ densities from the response tensor $\chi_{nn}(\mathbf{k})$, at temperature $\mathcal{T}=0.399$. Light blue and light red respectively denote $\chi_{nn}^{L}(k)$ and $\chi_{nn}^{T}(k)$, obtained from collating values over all orientations of $\mathbf{k}$; the spread in values at $k>0$ is due to our anisotropic wavenumber cutoff. Blue and red lines mark corresponding low-$k$ fits, with circles marking the extrapolated $k=0$ values. Purple and green squares mark the expected long-wavelength limits, as labeled. (b) Fits to $\chi_{nn}^{T}(k)$ at various temperatures, as labeled. Colors are as in (a). (c) Mass superfluid density $\rho_{nn}$ as a function of temperature, computed with multiple system sizes, as labeled. Inset shows the number of effectively uncorrelated samples $\tilde{\mathcal{N}}_{s}$ of the normal density $\varrho_{nn}$, as a function of temperature.}\label{fig:sfcomps}
\end{figure}

We compute the superfluid densities (\ref{eq:sffromchi}) through implementation of the SPGPE model detailed in Secs. \ref{sec:sgpe} and \ref{Sec:implementation2D}. An example, wherein we compute the mass superfluid density $\rho_{nn}$ of a gas with planar magnetized ($\lambda=0$) ground state, is shown in Fig.~\ref{fig:sfcomps} (a). At all temperatures $0.05<\mathcal{T}<0.5$ we find the long-wavelength limit of $\chi_{ij}^{T}(k)$ is in agreement with Eq.~(\ref{eq:varrho}). Alternatively, at temperatures away from any superfluid transition, we find $\varrho_{ij}$ may be reliably computed by extrapolating the finite $k$ behaviour of $\chi_{ij}^{T}(k)$ via a concave-down quadratic fit [see Fig.~\ref{fig:sfcomps} (b)]. Contrarily, within the $C$-field treatment we find the long-wavelength limit of $\chi_{ij}^{L}(k)$ does not agree with the expectation $\langle\Psi^{\dagger}\Psi\rangle$ at temperatures $\mathcal{T}>0$. However, it may instead be computed reliably by extrapolation of $\chi_{ij}^{L}(k)$ using a linear fit. The resulting mass superfluid density $\rho_{nn}=n_{nn}-\varrho_{nn}$ is shown in Fig.~\ref{fig:sfcomps}~(c). A mass BKT transition is evident at temperature $\mathcal{T}_{n}\approx 0.4$. System-size dependence is observed only above this transition temperature.

When computing equilibrium expectation values via time averaging (\ref{eq:timeav}), it is important to note that the samples $\mathcal{O}_{i}$ of an observable $\mathcal{O}$ obtained through the SPGPE (\ref{eq:sgpe}) will necessarily exhibit some degree of autocorrelation, dependent on the sampling interval. This manifests as an increase in the statistical error of the sampling $\Delta^{2} = \langle\left[\bar{\mathcal{O}}-\langle\mathcal{O}\rangle\right]^{2}\rangle$, given as
\begin{equation}
	\Delta^{2} = \frac{\text{Var}(\mathcal{O})}{\mathcal{N}_{s}}+\frac{2}{\mathcal{N}_{s}}\sum_{j=1}^{\mathcal{N}_{s}-1}\left[\langle\mathcal{O}_{1}\mathcal{O}_{1+j}\rangle-\langle\mathcal{O}\rangle^{2}\right],
\end{equation}
where $\text{Var}(\mathcal{O})$ denotes the variance of $\mathcal{O}$. In the absence of sample autocorrelation this reduces to $\Delta^{2}=\text{Var}(\mathcal{O})/\mathcal{N}_{s}$, motivating the definition of an effective number of uncorrelated samples $\tilde{\mathcal{N}}_{s}$, as
\begin{equation}
	\Delta^{2} = \frac{\text{Var}(\mathcal{O})}{\tilde{\mathcal{N}}_{s}}.\label{eq:Neff}
\end{equation}
We compute $\tilde{\mathcal{N}}_{s}$ via the algorithm presented in Ref.~\cite{ambegaokar_estimating_2010}. Beginning with the original sequence of samples $\mathcal{O}_{i}^{(0)}=\mathcal{O}_{i}$ we iteratively construct new sequences via the procedure
\begin{equation}
	\mathcal{O}_{i}^{(l)} = \frac{1}{2}\left[\mathcal{O}_{2i-1}^{(l-1)}+\mathcal{O}_{2i}^{(l-1)}\right],
\end{equation}
with $i \in\left\{1,2,\ldots,2^{-l}\mathcal{N}_{s}\right\}$. As $l$ increases the samples $\mathcal{O}_{i}^{l}$ maintain the same mean, yet exhibit a decreased degree of autocorrelation. We therefore expect the quantity
\begin{equation}
	\Delta_{l}^{2} = \frac{1}{\left(2^{-l}\mathcal{N}_{s}\right)^{2}}\sum_{i=1}^{2^{-l}\mathcal{N}_{s}}\left[\mathcal{O}_{i}^{(l)}-\bar{\mathcal{O}}^{(l)}\right]^{2}
\end{equation}
to increase with increasing $l$, ultimately saturating at $\Delta^{2}$. If so, comparison with Eq.~(\ref{eq:Neff}) allows for computation of $\tilde{\mathcal{N}}_{s}$. If saturation of $\Delta_{l}^{2}$ does not occur, sample autocorrelations remain significant, and a longer sampling time is required for estimation of $\Delta^{2}$. The results presented in Fig.~\ref{fig:sfcomps} were obtained by taking $\mathcal{N}_{s}=2.5\times 10^{5}$ steady-state samples $\Psi(t_{i})$ at uniform time intervals of $10\hbar/\mu$. As an example, the temperature dependence of the number $\tilde{\mathcal{N}}_{s}$ of effectively uncorrelated samples of the normal density $\varrho_{nn}$ (\ref{eq:varrho}) is inset to Fig.~\ref{fig:sfcomps} (c). Notably, we find that within the superfluid regime, this decreases with decreasing temperature. Furthermore, immediately above the mass superfluid transition temperature $\mathcal{T}_{n}$, the effective sample number drops as low as $\tilde{\mathcal{N}}_{s}\approx 20,000$. We note that uncorrelated samples may be alternatively computed by evaluating an ensemble of trajectories of the spin-1 SGPE, each beginning from a randomized initial state. However, the approach employed in this paper has the benefit of improved computational efficiency, permitted as the sample autocorrelation times are much smaller than the thermalization time.

\section{Superfluid phases of a 2D spin-1 ferromagnetic Bose gas}\label{sec:phases}

\begin{figure}
	\includegraphics[width=\linewidth]{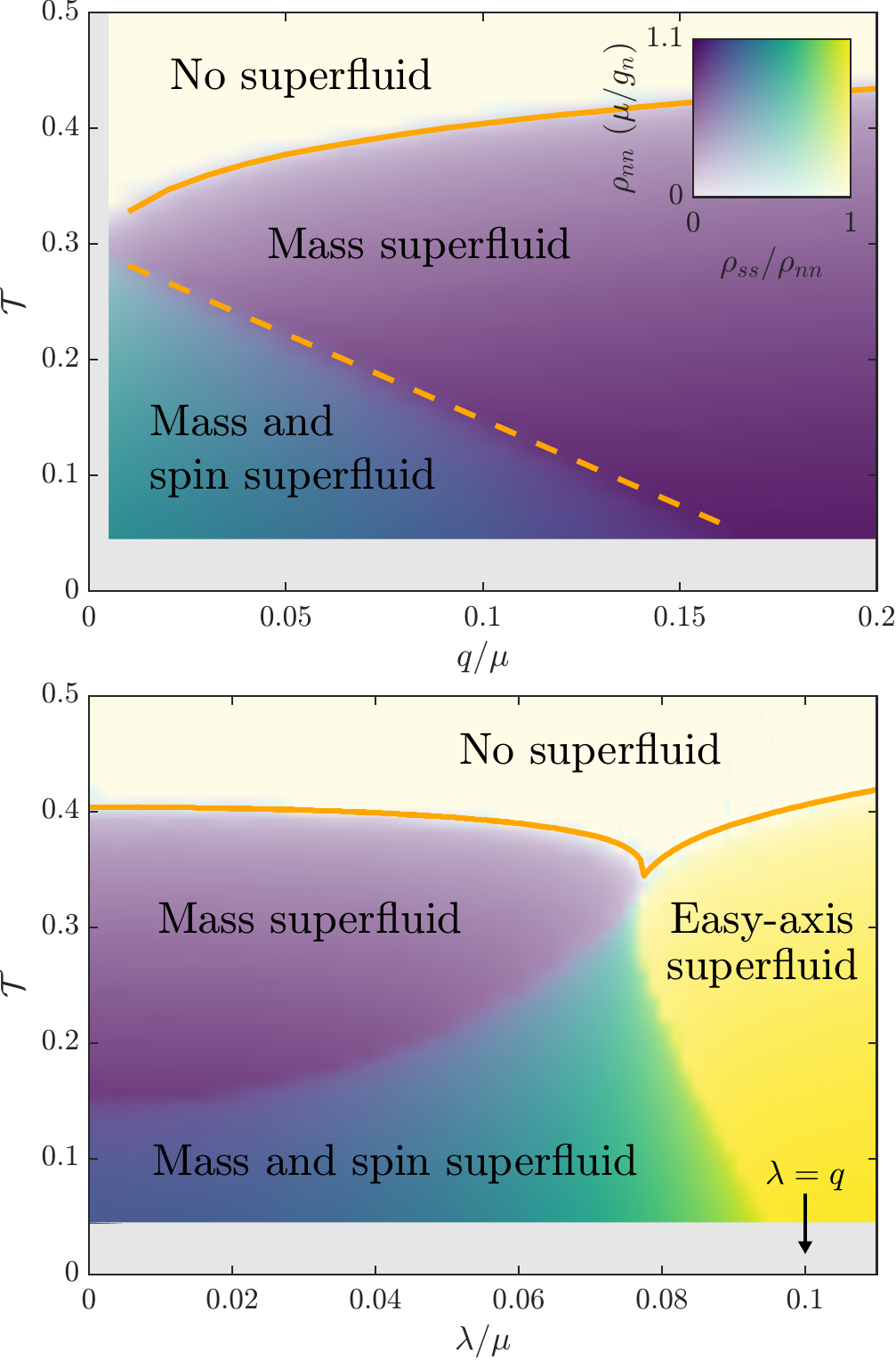}
	\caption{Superfluid phases of a spin-1 gas with BA ground state. Top: Quadratic Zeeman $q$ dependence with $\lambda=0$. Solid orange line denotes the mass transition temperature estimate $\mathcal{T}_{0}$ (\ref{eq:Tdens}). Dashed orange line is a linear fit to the spin transition temperature $\mathcal{T}_{s}$, see~Eq.~(\ref{eq:lindep}). Colorbar is inset. Bottom: Magnetic potential $\lambda$ dependence with $q=0.1\mu$. Solid orange line denotes the mass transition temperature estimate, $\text{max}\left\{\mathcal{T}_{0},\mathcal{T}_{1}\right\}$.}\label{fig:summary}
\end{figure}

In this section we apply the spin-1 SPGPE and the previously described superfluid measures to characterize the behavior of the  2D broken-axisymmetric Bose gas: In the first subsection we summarize the superfluid properties, which were previously detailed in Refs.~\cite{PhysRevResearch.5.L012045,PhysRevA.110.013311}. In the second subsection, we add some useful analytic characterization of these results that we have developed from the BKT theory for a scalar Bose gas \cite{PhysRevLett.87.270402}. In the third subsection, we identify the proliferation of free vortices in the spin-components and transverse magnetization, and relate this to the system superfluidity. In the fourth subsection we extend our results to provide a finite-temperature generalization of the phase diagram in Fig.~\ref{fig:groundstates}.

\subsection{Superfluidity in the  2D broken-axisymmetric Bose gas}
As discussed in Sec.~\ref{Sec:BA}, a spin-1 Bose gas with ferromagnetic interactions subject to a small positive quadratic Zeeman shift energetically favours the formation of spin order. As such, a 3D spin-1 gas in this regime is able to break two continuous symmetries of the Hamiltonian: The $\mathrm{U}(1)$ gauge symmetry associated with the global phase, and the $\mathrm{SO}(2)$ rotational symmetry associated with the transverse magnetization. The breaking of these symmetries manifests in the ability of this fluid to support superflow of both mass and spin currents. In the 2D system, thermal fluctuations preclude the formation of long-range order, however the aforementioned superfluidities can still arise via BKT transitions, where the associated superfluid densities describe the system rigidity against twists in the associated fields. This can be quantified in SGPE calculations using the formalism we developed in Sec.~\ref{sec:superfluidity}.

Here we focus on the superfluid properties of a uniform 2D system, considering equilibrium states within a range of $q$ and $T$ values. The simulation size varies with temperature according to the procedure outlined in Sec.~\ref{Sec:Cregion}, so as to define a physically consistent $C$-region. In each case we extract the superfluid densities from the long-wavelength limits of the response tensors $\chi_{ij}$ (cf.~Fig.~\ref{fig:sfcomps}), which are computed through sampling of the equilibrium states. Here we take the interaction parameters $|\tilde{g}_s|/\tilde{g}_n=0.1$, as previously used in this paper. This ratio is between the the rather small value realized in \textsuperscript{87}Rb~\cite{PhysRevLett.87.010404} ($|\tilde{g}_s|/\tilde{g}_n\sim10^{-2}$) and the large value obtained in \textsuperscript{7}Li~\cite{huh2020} ($|\tilde{g}_s|/\tilde{g}_n\sim0.5$). While not shown here, we have additionally considered $|\tilde{g}_{s}|/\tilde{g}_{n}=0.5$, and found the results to be qualitatively similar.
	
The results of these calculations, summarizing the superfluid behavior of 2D broken-axisymmetric Bose gases, are given in Fig.~\ref{fig:summary}. First consider the broken-axisymmetric gas with $\lambda=0$. The mass BKT transition, earlier demonstrated in Fig.~\ref{fig:sfcomps}, is analogous to that observed in a single-component Bose gas. In particular, mass superfluidity $\rho_{nn}$ emerges at temperature $\mathcal{T}_{n}$ driven by the binding of mass vortex-antivortex pairs. Within the mass-superfluid regime, the reduction of $\rho_{nn}$ from its zero-temperature value $n$ is due to thermally excited long-wavelength phase fluctuations. The breaking of $\mathrm{SO}(2)$ spin-rotational symmetry by the planar-magnetized ground state manifests in the existence of an additional BKT transition. Here spin superfluidity $\rho_{ss}$ emerges at temperature $\mathcal{T}_{s}$, driven by the binding of spin vortex-antivortex pairs. Within the spin-superfluid regime, the reduction of $\rho_{ss}$ from its zero temperature value $\langle \Psi^{\dagger}f_{z}^{2}\Psi\rangle$ is due to thermally excited long-wavelength fluctuations in the spin orientation $\alpha$. 

\begin{figure}
	\includegraphics[width=\linewidth]{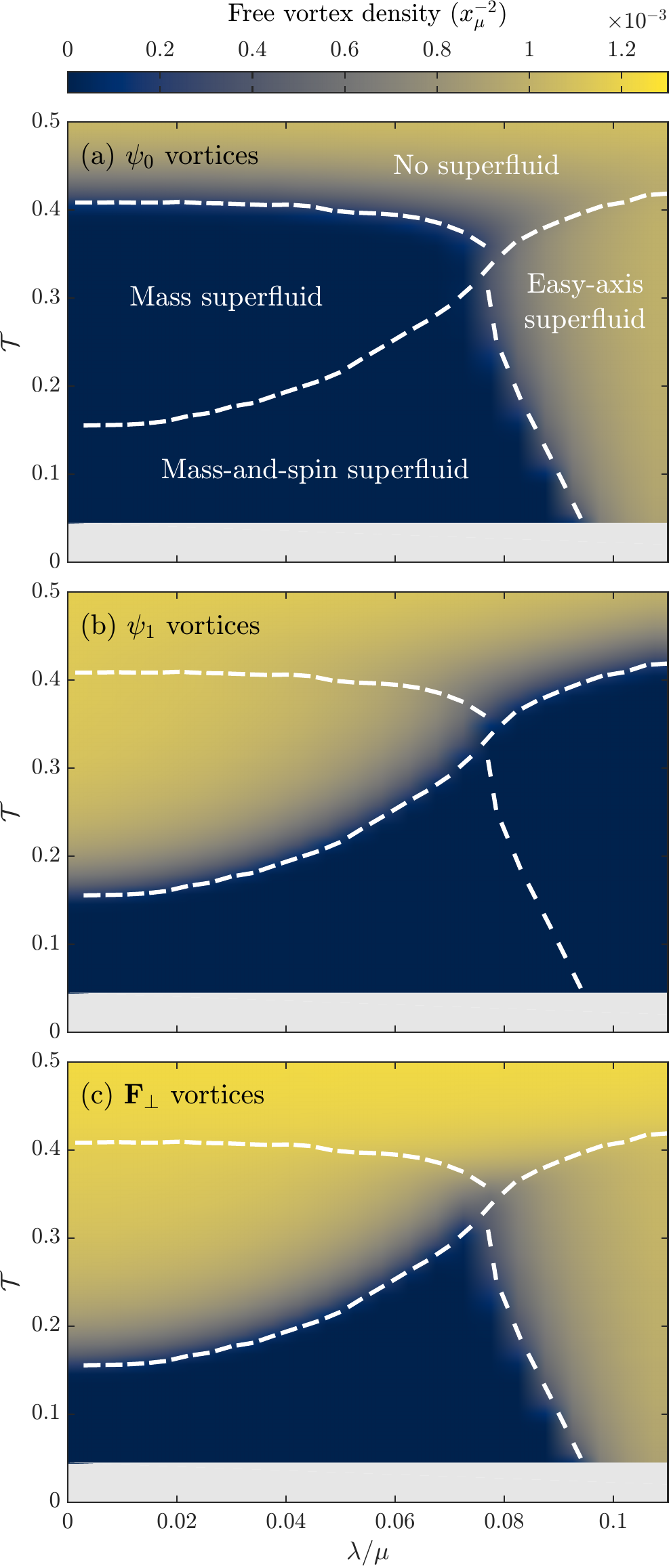}
	\caption{Equilibrium density of free vortices in the fields (a) $\psi_{0}$, (b) $\psi_{1}$, and (c) $\mathbf{F}_{\perp}$, within a spin-1 gas with BA ground state. Each panel shows magnetic potential $\lambda$ and temperature $\mathcal{T}$ dependence, with $q=0.1\mu$. Dashed white lines indicate approximate boundaries of the three superfluid phases, see Fig.~\ref{fig:summary}.}\label{fig:vortices}
\end{figure}

\subsection{Component density based superfluid transition predictions}
Numerical results similar to those in Fig.~\ref{fig:summary} were presented in our previous work. Here we additionally show that we can analytically quantify the phase transitions by utilizing the single-component Bose gas result~\cite{PhysRevLett.87.270402}  
\begin{equation}
	n  = \frac{Mk_{B}T}{2\pi\hbar^{2}}\ln\left(\frac{380\hbar^{2}}{Mg}\right).\label{eq:Tn}
\end{equation}
This result provides a relationship between the density $n$, the coupling constant $g$, and temperature $T$ at the critical point. This can be contrasted with the Nelson-Kosterlitz result $\rho  =  2Mk_{B}T/\pi\hbar^{2}$ for the superfluid density $\rho$ at the critical point. Motivated by Eq.~(\ref{eq:Tn}), we define a (dimensionless) temperature $\mathcal{T}_{m}$ for each component $m\in\left\{1,0,-1\right\}$ of the spin-1 system as:
\begin{equation}
	n_{m}(\mathcal{T}_{m}) = \frac{\mu\mathcal{T}_{m}}{2\pi g_{n}}\ln\left(\frac{380}{\tilde{g}_{n}}\right),\label{eq:Tdens}
\end{equation} 
with $n_{m} = \langle \psi_{m}^{*}\psi_{m}\rangle $ the component densities. At all values of $q$ we find that $\mathcal{T}_{n}\approx \mathcal{T}_{0}$, i.e. the quadratic Zeeman dependence of the mass transition temperature may be attributed to variation in the $m=0$ spin component density. Regarding $\mathcal{T}_{s}$, we find a linear fit
\begin{equation}
	\mathcal{T}_{s}\propto 1-\frac{q/\mu}{|g_{s}|/g_{n}} \label{eq:lindep}
\end{equation}
is applicable at all $0<q<2\mu|g_{s}|/g_{n}$. Such a linear dependence follows from the ground-state densities of the $m=\pm 1$ spin components, $n_{\pm 1}(\mathcal{T}=0) = n(\mathcal{T}=0)/4-q/8|g_{s}|$ (along the dashed line in Fig.~\ref{fig:summary} the $m=1$ spin component density is within $20\%$ of its ground state value).

With nonzero magnetic potential $\lambda>0$ we again observe mass and spin BKT transitions, however the nature of these is modified by the presence of axial magnetization $F_{z}>0$. In particular, with $\lambda>0$ mass and spin superflows are no longer independent, reflected by nonzero values of $\rho_{ns}$. This is exemplified at large $\lambda\sim q$, where the mass and spin superfluid densities are identical $\rho_{nn}=\rho_{ss}=\rho_{ns}$. Within this regime, despite nonzero population of all three spin components, the superfluid properties are similar to that of a gas with easy-axis ground state. As such, we term this the easy-axis superfluid phase [yellow region in Fig.~\ref{fig:summary}]. At all values of $\lambda\geq0$ we find the mass transition temperature is well estimated as $\mathcal{T}_{n}\approx\text{max}\{\mathcal{T}_{0},\mathcal{T}_{1}\}$.

\subsection{Role of vortices in phase transitions}
Transitions to the three superfluid phases are again driven by the binding of vortex-antivortex pairs. With $\lambda>0$, the vortices of primary interest are those in the fields $\psi_{0}$, $\mathbf{F}_{\perp}$, and $\psi_{1}$ (at $\lambda =0$, vortices in $\psi_{0}$ and $\mathbf{F}_{\perp}$ are respectively identified as mass and spin vortices, while at $\lambda\gg q$ vortices in $\psi_{1}$ are equivalently identified as mass or spin vortices). We quantify vortex-antivortex binding by computing the equilibrium density of free vortices in each of the aforementioned fields. In doing so, we count the number of free defects in a given steady-state sample $\Psi$ by convolving the desired complex field (here we take $\mathbf{F}_{\perp}\to F_{x}+\mathrm{i}F_{y}$) with a Gaussian filter, before locating points around which its phase winds by $\pm 2\pi$. The convolution acts to remove vortex-antivortex pairs on length scales smaller than a specified filter width; we take this width to be $5\sqrt{5}x_{\mu}$, motivated by the size of a zero-temperature spin vortex~\cite{PhysRevResearch.3.013154}. Results are shown in Fig.~\ref{fig:vortices}. As in the un-magnetized case, the mass superfluid phase is associated with the absence of free $\psi_{0}$ vortices, while the mass-and-spin superfluid phase is associated with the additional absence of free $\mathbf{F}_{\perp}$ vortices. Within the easy-axis superfluid phase, we observe the absence of free vortices only in the field $\psi_{1}$.

\subsection{Finite-temperature $(q,\lambda)$-phase diagrams}

\begin{figure}
	\includegraphics[width=\linewidth]{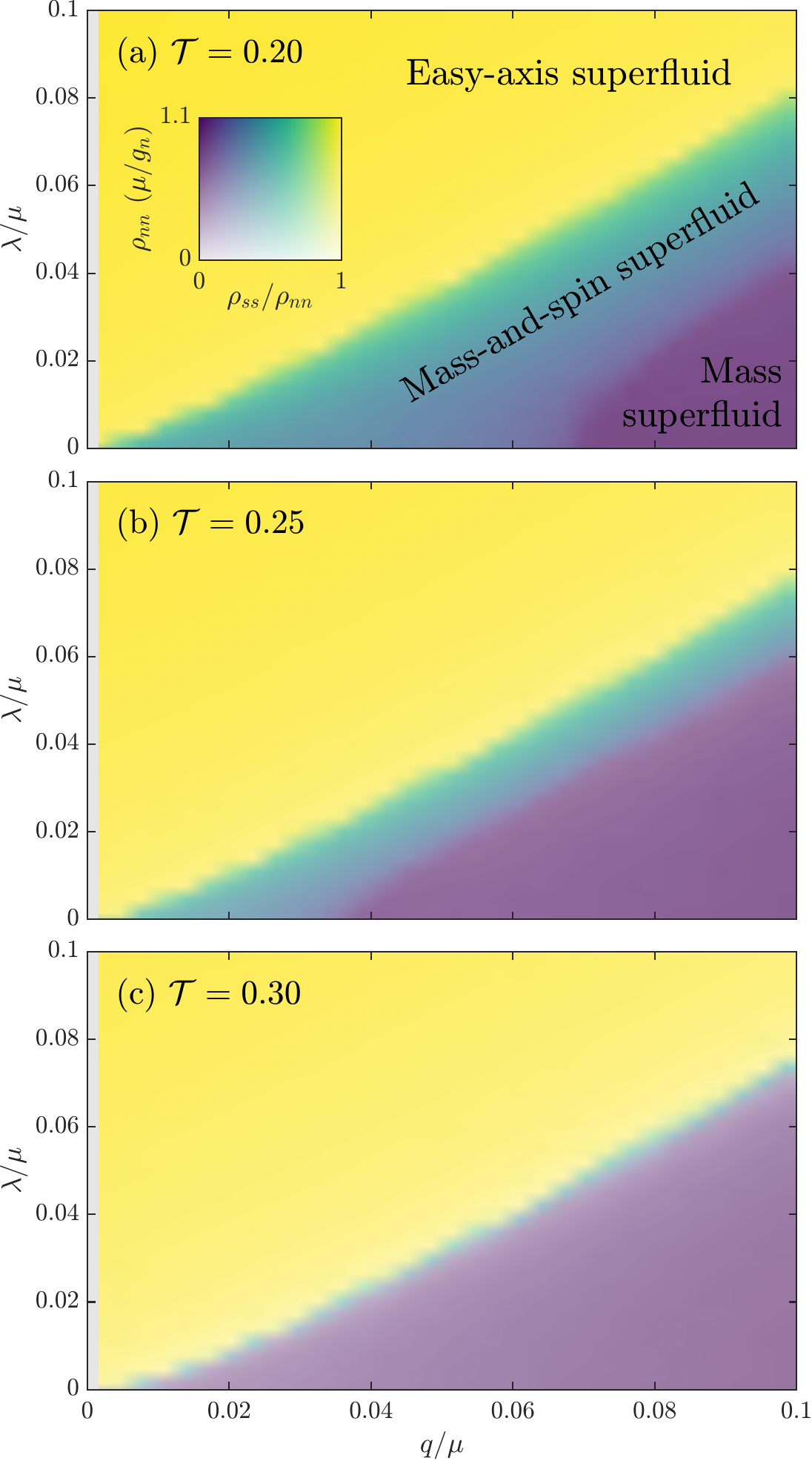}
	\caption{Superfluid phases of the spin-1 Bose gas with ferromagnetic interactions at temperatures (a) $\mathcal{T}=0.20$, (b) $\mathcal{T}=0.25$, and (c) $\mathcal{T}=0.30$. Colorbar is inset in panel (a).}\label{fig:Tdiagrams}
\end{figure}

For a final set of results, we construct finite-temperature phase diagrams detailing the dependence of the system superfluid densities on the quadratic Zeeman energy and magnetic potential. At zero temperature, the spin-1 gas is characterized by its ground state, according to a diagram such as that presented in Fig.~\ref{fig:groundstates}. At nonzero temperature, the spin-1 gas may be alternatively characterized by its superfluid properties; the phase diagrams presented in this section may thus be considered finite-temperature generalizations of Fig.~\ref{fig:groundstates}. In particular, we consider the mass superfluid regime a generalization of the polar phase, the mass-and-spin superfluid regime a generalization of the broken-axisymmetric phase, and the easy-axis superfluid regime a generalization of the easy-axis phase. The resulting generalizations of Fig.~\ref{fig:groundstates} are depicted in Fig.~\ref{fig:Tdiagrams}.

First consider $\mathcal{T}=0.2$. While the structure observed in Fig.~\ref{fig:groundstates} is preserved, the parameter regime of the mass and spin superfluid phase has been greatly reduced. This may be understood from the results of Fig.~\ref{fig:summary}. First, the $\lambda=0$ transition from the mass superfluid phase to the mass-and-spin superfluid phase has decreased from the zero temperature value of $q = 0.2\mu$ to $q\approx 0.07\mu$, in accordance with the relation (\ref{eq:lindep}) observed in Fig.~\ref{fig:summary} (a). Second, the transition from the mass-and-spin superfluid phase to the easy-axis superfluid phase has been reduced from $\lambda=q$ to $\lambda<q$, as seen in Fig.~\ref{fig:summary} (b). Upon further increases in temperature the parameter regime of the mass-and-spin superfluid phase continues to decrease, until it vanishes at $\mathcal{T}\approx 0.3$.

\section{Conclusion and outlook}\label{sec:concs}
 
 In this paper, we developed and applied a SPGPE model for simulating finite-temperature physics in a spin-1 Bose gas. This framework extends c-field methods to spinor systems, enabling a range of studies of equilibrium and dynamical properties. 
 The simulated $C$-region of the gas consists of the single-particle modes with significant occupation, being explicitly defined with use of a low-energy projector. For the applications in this paper, we have focused on the equilibrium physics of a 2D spin-1 Bose gas with ferromagnetic interactions. A valid physical treatment of this strongly fluctuating system requires that the $C$-region, in addition to being comprised of significantly occupied modes, also contains all strongly coupled modes. Provided the temperature is sufficiently high, $k_{B}T\gtrsim\mu$, these conditions can be simultaneously satisfied through implementation of a projector enforcing a suitable high-momentum cutoff.
 
 The superfluid phases of a spin-1 Bose gas were identified, and quantified using both generalized momentum fluctuations and the long-wavelength limits of current-current correlation functions. We found three distinct superfluid phases: a mass superfluid phase, an easy-axis superfluid phase, and a mass-and-spin superfluid phase—finite-temperature generalizations of the polar, easy-axis, and broken-axisymmetric ground states. Our results for a finite temperature phase diagram show that as temperature increases, the stability region of the mass-and-spin superfluid phase shrinks and eventually vanishes. We have additionally presented alternative methods to identify the phase transitions, potentially of use for experiments: First, we have shown that one can utilize a general result for the BKT transition in a scalar 2D Bose gas to predict the mass and spin superfluid transition temperatures from the component densities. Second, we have analyzed the unbinding of vortices in both the field components and the transverse magnetization, showing that this provides a clear indication of the distinct phases.

The spin-1 SPGPE implementation introduced in this work, along with our characterization of equilibrium phases, establishes a foundation for future studies of finite-temperature spinor Bose gases. 
%This model can be extended to investigate the effects of spatial inhomogeneity, where variations in density or magnetization may lead to the coexistence of multiple superfluid phases, offering insights into finite-temperature topological interface physics. Additionally, this framework enables the study of nonequilibrium dynamics, including temperature, quadratic Zeeman energy, and magnetic potential quenches. Another intriguing direction is the exploration of dimensional crossover by extending the model to a three-dimensional gas with strong harmonic confinement along the axial direction. In scalar Bose gases, such a crossover leads to a transition from BKT physics to Bose-Einstein condensation, and the interplay of density and spin degrees of freedom in spin-1 systems may present novel crossover behavior.
%Together, the SPGPE model detailed in Sec.~\ref{sec:sgpe}, and the characterization of the spin-1 Bose gas equilibrium superfluid properties presented in Sec.~\ref{sec:phases}, provide a foundation upon which to perform further study of the finite-temperature spin-1 Bose gas. 
We comment on a few directions which could be explored using the approach discussed here with minor changes.
First, an interesting direction would be to implement inhomogeneity into the description. This could be the result of external trapping (leading to density variations) or in the Zeeman fields leading to spatial variations in either the axial magnetization or the expected equilibrium phase. This may result in the realization of domains with distinct phases, potentially elucidating the effect of finite temperature on topological interface physics in the spin-1 Bose gas (see~\cite{PhysRevLett.109.015302,PhysRevA.87.033617,OBorgh_2014}). A second avenue of extension for this theory is the study of quench dynamics induced by a sudden change of the reservoir parameters (e.g.~a sudden change in $T$, $\mu$ or $\lambda$, see~\cite{PhysRevA.84.063625}), or of the quadratic Zeeman shift $q$. This will complement the existing body of work on zero-temperature quenches in the spin-1 system induced by changes in external fields (e.g.~see~\cite{sadler_spontaneous_2006,PhysRevA.84.023606,PhysRevLett.99.130402,PhysRevLett.98.160404,PhysRevA.76.043613,PhysRevA.75.013621,PhysRevA.96.013602,PhysRevLett.116.025301,prufer_observation_2018,PhysRevA.94.023608,PhysRevA.99.033611,PhysRevA.95.023616,PhysRevA.100.033603,PhysRevLett.119.255301,PhysRevA.98.063618,PhysRevLett.122.173001,PhysRevLett.122.170404,10.21468/SciPostPhys.7.3.029, PhysRevLett.131.183402,PhysRevLett.131.183402}). 
In relation to the 2D superfluid transitions of the spinor system, it would be of interest to consider the 2D-3D crossover which occurs as the $z$-confinement changes. In a scalar Bose gas such a dimensional crossover is predicted to result in a continuous shift in the nature of the observed phase transition, from BKT to Bose-Einstein condensation~\cite{PhysRevResearch.4.033130} (also see \cite{Hadzibabic_2008,chomaz_emergence_2015,PhysRevLett.114.255302}). The spin-1 system considered here may provide a fruitful extension of this concept, owing to the separation between density and spin healing lengths (e.g.,~the two differ by an order of magnitude in \textsuperscript{87}Rb~\cite{PhysRevLett.88.093201}).

\section{Acknowledgements}

The authors would like to acknowledge the contributions of L.~A.~Williamson, Xiaoquan Yu and Andrew.~J.~Groszek in prior work focused on the BKT physics of the spin-1 ferromagnetic system.

\end{document}